\documentstyle[amsfonts]{article}
\def\cM{{\cal M}_\rho^6}
\def\cm{{\cal M}_\rho^4}
\def\SO{SO^{\uparrow}(3,2)}
\def\pa{\partial}
\def\cG{{\cal G}}
\def\ob{\stackrel{\circ}{b}}
\def\oy{\stackrel{\circ}{y}}
\def\cF{{\cal F}}
\def\cC{{\cal C}}
\def\dx{{\dot x}}
\def\dz{{\dot z}}
\def\dbz{\dot{\bar z}}
\def\La{\Lambda}
\def\dF{\dot{\cal F}}
\def\dxi{{\dot\xi}}
\title{Anti-de Sitter spinning particle\\
and two-sphere}
\author{S. M. Kuzenko$^{1,2}$, S. L. Lyakhovich$^1$, \\
A. Yu. Segal$^1$, and A. A. Sharapov$^1$}
\date{$^1$Department of Physics, Tomsk State University,\\
Tomsk 634050, Russia\\
e-mail: SLL@fftgu.tomsk.su\\
$^2$Institute for Theoretical Physics, Hannover University,\\
2 Appelstr., D-3000 Hannover 1, Germany}
\begin{document}
\Large
\maketitle
\begin{abstract}
We propose the action of $d=4$ Anti-de Sitter (AdS) spinning particle
with arbitrary fixed quantum numbers. Regardless of the spin value, the
configuration space is $\cm\times S^2$ where $\cm$
is $d=4$ AdS space, and two-dimensional sphere $S^2$ corresponds to
spinning degrees of freedom. The model is an AdS counterpart of the
massive spinning particle in the Minkowski space proposed earlier.
Being AdS-invariant, the model possesses two gauge symmetries implying
identical conservation law for AdS-counterparts of mass and spin. Two
equivalent forms of the action functional, minimal and manifestly
covariant, are given.
\end{abstract}
\newpage
\section{Introduction}

We suggest a new model of a spinning particle which propagates in $d=4$
Anti-de Sitter space and has arbitrary fixed quantum numbers. The model
is an AdS counterpart of the massive spinning particle theory in the
Minkowski space proposed in Ref. [1].

A consistency of the interaction with external fields (including
gravity) has always been a problem for a higher spin (super)particle
theory. In this connection the model being studied is of special
interest as a simplest example of a consistent spinning particle theory
in the curved space which could probably be treated as a suitable
background for perturbative interaction switching on. In relation to
this topic it is pertinent to note that just the AdS space appears to
be an admissible vacuum for interacting higher spin fields (including
gravity) [2].

Let us discuss in outline the starting points of the model's
construction. The configuration manifold is chosen\footnote{\normalsize{This
choice of the configuration manifold differs from the other ones, being
usually used for spinning particles (e.g. see [3--7]), in that it does
not depend on the spin value.}} as $\cM=\cm\times S^2$, where
$\cm$ is $d=4$ Anti-de Sitter space ($\rho <0$ is the
curvature) and $S^2$ is two-sphere. The case of $\rho=0$ (i.e.,
${\cal M}^4_0\equiv {\Bbb R}^{3,1}$) corresponds to the model studied in Ref.
[1]. It turns out that $\cM$ can be endowed with a structure of a
homogeneous space for AdS group (see Sec. 2). Thus, $\cM$ is able to
serve as an arena for some AdS-invariant dynamics.

There is a number of AdS-invariant functionals of world-line on
$\cM$, and each of them can seemingly be treated as an appropriate
action for the spinning particle. However, we are going to show that
the action functional will be unambigously determined if an identical
conservation law is required to hold for the AdS-counterparts of spin
and mass\footnote{\normalsize{It means, in fact, that the AdS particle
spin and mass should appear to coincide identically (i.e., off shell)
with some numerical parameters entering the action desired.}}. Thus, the
basic selection principle is that the action should possess two gauge
symmetries being provided the pair of N\"oether identities to appear.
{}From the standpoint of Hamiltonian formalism this principle means that a
pair of the AdS-invariant first-class constraints should be imposed
onto the cotangent bundle of $\cM$ to extract physically contentable
degrees of freedom. On the other hand, it turns out that just the
theory with two gauge invariances in $\cM$ has the proper number of the
physical degrees of freedom being characterized the spinning particle:
4 = 3(positions) + 1(spin). The mentioned properties of the model are
shown to cause the spinning particle theory quantization to give the
irreducible representation of AdS group.

The letter is organized as follows. In Sec. 2 we study an AdS-covariant
description for the configuration manifold and show that $\cM$ is a
homogeneous transformation space for AdS group. In Sec. 3, we derive
the model action functional in an explicitly AdS-covariant manner and
discuss its local symmetries both in the first- and the second-order
formalism. In Sec. 4, we consider the model's description in terms of
inner $\cM$ geometry. It shows also that the model can be treated as a
``minimal covariant extention'' of its flat-space counterpart [1]. We
also consider in the Section some obstructions to straightforward
generalization of the model to the case of arbitrary curved background.
The Conclusion includes discussion of the results and some
perspectives.

\section{Covariant realizations for the configuration space}

We start with describing two covariant realizations for the
configuration space $\cM=\cm\times S^2$, where
$\cm$ presents itself an ordinary Anti-de Sitter space,
$S^2$ is two-dimensional sphere. It is useful for us to treat
$\cm$ as a hyperboloid embedded into a five-dimensional
pseudo-Euclidean space ${\Bbb R}^{3,2}$, with coordinates $y^A$, $A=5,
0, 1, 2, 3$, and defined by
$$
\eta_{AB}y^Ay^B=-R^2, \qquad \eta_{AB}=(--+++),
\eqno{(1)}$$
$\rho =-R^{-2}$ is the curvature of the AdS space. There is no problem,
however, to extend subsequent results to the case when $\cm$ stands for
the universal covering space of the hyperboloid.

Similarly to $\cm$, $\cM$ can be endowed with the structure of a
homogeneous transformation space for an AdS group chosen below to be
the connected component of unit in $O(3,2)$ and denoted by
$\SO$\footnote{\normalsize{The elements of $\SO$ are specified
by the conditions that their diagonal $2\times 2$ and $3\times 3$
submatrices, numbering by indices $\overline{5,0}$ and $\overline{1,2,3}$
respectively, have positive determinants.}}. In order to explain this
statement, let us consider the tangent bundle $T(\cm)$ that will be
parametrized by 5-vector variables $(y^A,b^A)$ under the constraints
$$
y^Ay_A=-R^2,
\eqno{(2.a)}$$
$$
y^Ab_A=0.
\eqno{(2.b)}$$

The latter requirement simply expresses the fact that $b^A\pa/\pa y^A$
is a tangent vector to a point $y\in\cm$. The $O(3,2)$-invariant
subbundle $\tilde T(\cm)$ of non-zero light-like tangent vectors is
embedded into $T(\cm)$:
$$
b^Ab_A=0,
\eqno{(3.a)}$$
$$
\{b^A\}\neq 0.
\eqno{(3.b)}$$
It turns out that $\cM$ can be identified with the factor-space of
$\tilde T(\cm)$ with respect to the equivalence relation
$$
b^A\sim\lambda b^A, \qquad \forall\lambda\in{\Bbb R}\setminus\{0\}.
\eqno{(4)}$$
Really, there always exists a smooth mapping
$$
{\cal G}: \cm\to \SO
\eqno{(5)}$$
such that ${\cal G}(y)$ moves a point $(y,b)$ at $\tilde T(\cm)$ to
$(\oy,\ob)$ having the form
$$
\oy{}^A = {{\cal G}^A}_B(y)y^B=(R,0,0,0,0)
\eqno{(6)}$$
and
$$
\ob{}^A = {{\cal G}^A}_B(y)b^B=(0,u^a) \qquad
a=0,1,2,3,
\eqno{(7.a)}$$
$$
\eta_{ab}u^au^b=0,
\eqno{(7.b)}$$
$$
\{u^a\}\neq 0.
\eqno{(7.c)}$$
For example, one can choose
$$
{\cal G}(y)=\left(\begin{array}{ccc}\begin{array}{cc}
\frac{\displaystyle{y^5}}{\displaystyle{R}} & \frac{\displaystyle{y^0}}
{\displaystyle{R}}\\
-~\frac{\displaystyle{y^0}}{\displaystyle\rho} & \frac{\displaystyle
y^5}{\displaystyle\rho}\end{array} &\stackrel{\vdots}{\vdots}&
\begin{array}{rrr}
-~\frac{\displaystyle y^1}{\displaystyle R} & -~\frac{\displaystyle
y^2}{\displaystyle R} & -~\frac{\displaystyle y^3}{\displaystyle R}\\
{\displaystyle 0} & {\displaystyle 0} & {\displaystyle 0}\end{array}\\
\dotfill &\vdots& \dotfill\\
\begin{array}[c]{cc}-~\frac{\displaystyle y^1y^5}{\displaystyle R\rho}
& -~\frac{\displaystyle y^1y^0}{\displaystyle R\rho}\\
-~\frac{\displaystyle y^2y^5}{\displaystyle R\rho} &
-~\frac{\displaystyle y^2y^0}{\displaystyle R\rho}\\
-~\frac{\displaystyle y^3y^5}{\displaystyle R\rho} &
-~\frac{\displaystyle y^3y^0}{\displaystyle R\rho}\end{array} &\vdots&
{\displaystyle\delta^{ij}}+\frac{\displaystyle y^iy^j}{\displaystyle
R(\rho +R)}\end{array}\right),
\eqno{(8)}$$
where
$$
\rho =\sqrt{(y^5)^2+(y^0)^2}, \qquad i,j=1,2,3.
$$
{}From Eq. (7) we see that the fiber $\{(\oy,\ob)\}$ over $\oy$ in
$\tilde T(\cm)$ looks exactly like the punctured light-cone in
Minkowski space. Equivalence relation (4) proves to reduce the
light-cone to $S^2$. Now, since the AdS group brings any equivalent
points to equivalent ones, we conclude that $SO^{\uparrow}(3,2)$
naturally acts on the factor-space $\cm\times S^2$. Therefore, Eqs.
(2)-(4) present an AdS-covariant realization of $\cM$.

There exists some inherent arbitrariness in the choice of $\cal G$
defined by Eqs. (5) and (6). Such a mapping can be equally well
replaced by another one
$$
{{\cal G}'}^A{}_B(y)={\Lambda^A}_C(y){{\cal G}^C}_B(y),
\eqno{(9.a)}$$
where $\Lambda$ takes it values in the stability group of the marked
point $\oy$,
$$
{\Lambda^A}_B(y)\oy{}^B=\oy{}^A,
\eqno{(9.b)}$$
and has the general structure
$$
\begin{array}{l}
\Lambda : \cm\to\SO\\
{\Lambda^A}_B(y)=\left(\begin{array}{ccc} \displaystyle 1 & \vdots & 0\\
\dotfill & \vdots & \dotfill\\
\displaystyle 0 & \vdots & {\Lambda^a}_b(y)\end{array}\right), \qquad
{\Lambda^a}_b(y)\in SO^{\uparrow}(3,1).\end{array}
\eqno{(10)}$$
The set of all smooth mapping (10) forms an infinite-dimensional group
isomorphic to a local Lorentz group of the AdS space. This group acts
on $T(\cm)$ by the law
$$
(y,b)\longrightarrow\big(y,{\cal G}^{-1}(y)\Lambda(y){\cal G}(y)b\big),
\eqno{(11)}$$
$\cal G$ being a fixed solution of Eqs. (5), (6). As is obvious, the
local Lorentz group naturally acts on $\cM$.

For getting explicit action of $\SO$ on $\cM$, it appears useful from
the very beginning to replace the AdS-covariant parametrization
$(y^A,b^A)$ of $T(\cm)$ with a Lorentz-covariant one $(y^A,u^a)$, where
the 4-vector $u^a$ is related to $b^A$ as it is given by
Eq. (7.a). Given a group element $H\in\SO$, it moves $(y,b)$ to
$(Hy,Hb)$, hence $(y,u)$ to $(Hy,\Lambda_H(y)u)$, where
$$
\Lambda_H(y)\equiv{\cal G}(Hy)H{\cal G}^{-1}(y)
\eqno{(12)}$$
is a Lorentz transformation of the form (10). One readily finds
$$
\Lambda_{H_1}(y)\Lambda_{H_2}(y)=\Lambda_{H_1H_2}(y)
\eqno{(13)}$$
for arbitrary $H_1,H_2\in\SO$. We thus arrive at a nonlinear
representation of the AdS group. Now, the problem reduces to obtaining
action of the Lorentz group on $S^2$, what is well known and has been
described in detail in Ref. [1] in the convenient for our purposes
form. Also it will be given in Sec. 4.

\section{Derivation of the action functional}

Here we derive the model in an AdS-covariant way. The basic requirement
allowing us to choose the unique action functional is: the dynamical
content of the model on $\cM$ must be completely determined by
identical conservation of classical counterparts of two Casimir
operators of AdS group.

Let us consider the model's phase space with the coordinates $y^A,b^A$
and their canonically conjugated momenta $P_A,K_A$, subjected to the
following nonvanishing Poisson bracket relations:
$$
\{y^A,P_B\}=\{b^A,K_B\}=\delta^A_B.
\eqno{(14)}$$
AdS group acts on the phase-space functions via brackets (14) by the
following generators:
$$
L_{AB}=y_AP_B-y_BP_A+b_AK_B-b_BK_A.
\eqno{(15)}$$
The theory being constructed must contain the constraints (2), (3) as
well as the equivalence relation (4) to be well defined on $\cM$. Thus,
we impose the following AdS-invariant first-class constraints to
restrict an admissible dynamics of the model:
$$
T_{1,2,3,4}\approx 0,
\eqno{(16.a)}$$
where
$$
T_1=y_Ay^A+R^2, \qquad T_2=y_Ab^A, \qquad T_3=b_Ab^A,
\eqno{(16.b)}$$
$$
T_4=K_Ab^A.
\eqno{(16.c)}$$
The last constraint generates the equivalence relation (4) with respect
to the brackets (14). Being restricted to the surface (2), (3) the
classical counterparts of Casimir operators of the AdS group look as
$$
C_1=\frac{1}{2}L_{AB}L^{AB}|_{T_{1,2,3,4}=0}=R^2P_AP^A+(P_Ay^A)^2+
2(P_Ab^A)(K_Ay^A),
\eqno{(17.a)}$$
$$
C_2=\frac{1}{4}L_{AB}{L^B}_C{L^C}_DL^{DA}|_{T_{1,2,3,4}=0}=\qquad\qquad\qquad
$$
$$
\qquad\frac{1}{8}(L_{AB}L^{AB})^2-(P_Ab^A)^2[K_AK^AR^2+(K_Ay^A)^2].
\eqno{(17.b)}$$
Taking the proper account of the basic requirement formulated above in
this section and Eqs. (17.a-c) we introduce the two main first-class
constraints
$$
T_5=R^2P_AP^A+(P_Ay^A)^2+2(P_Ab^A)(K_Ay^A)+M\approx 0,
\eqno{(18.a)}$$
$$
T_6=(P_Ab^A)^2[K_AK^AR^2+(K_Ay^A)^2]+\delta\approx 0,
\eqno{(18.b)}$$
where $M$ and $\delta$ are some constants which are treated as
parameters of the model. It easily seen that
$$
C_1=-M, \qquad C_2=\frac{1}{2}M^2+\delta
\eqno{(19)}$$
on the total constrained surface.

Thus, the model postulated is characterized by six first-class
constraints: four of them $T_{1,2,3,4}$ are auxiliary ones (they reduce
the configuration space to $\cM$), while the two principal constraints
$T_5$, $T_6$ determine the dynamics on $\cM$ properly.

The first-order (Hamiltonian) action associated with these six
constraints is
$$
S=\int{\rm d}\tau \left(P_A\dot y^A+K_A\dot b{}^A-\sum_{i=1}^6
\frac{\nu_i}{2}T_i\right).
\eqno{(20)}$$
Here $\nu_i$ are auxiliary variables playing the role of Lagrange
multipliers to the first-class constraints $T_i$.

To bring this action to a second-order (Lagrange) form one must exclude
the momenta $P_A$, $K_A$ and the auxiliary constraints' multipliers
$\nu_{1,2,3,4}$ by making use of other equation of motion:
$$
\frac{\delta S}{\delta P_A}=0, \qquad \frac{\delta S}{\delta K_A}=0,
\qquad \frac{\delta S}{\delta\nu_{1,2,3,4}}=0.
\eqno{(21)}$$
It is easy to check that on the shell of Eq. (16, 18) the following
equalities hold:
$$
\dot y_Ab^A=\nu_5R^2(P_Ab^A),
\eqno{(22.a)}$$
$$
\dot y_A\dot y{}^A=\nu_5R^2\left(\nu_5(T_5-M)+2\nu_6(\delta-T_6)\right),
\eqno{(22.b)}$$
$$
\dot b_A\dot b^A=-\frac{1}{R^2}\frac{(\dot y_Ab^A)^2}{\nu^2_5}\left(\nu^2_5+
\frac{1}{2}\nu_6^2(\delta-T_6)\right),
\eqno{(22.c)}$$
$$
P_A\dot y^A+K_A\dot b^B=\nu_5(T_5-M)+2\nu_6(\delta-T_6),
\eqno{(22.d)}$$
Substituting these relations to Eq. (20) we are coming up with the
following action of the model:
$$
S_1=\int{\rm d}\tau\bigg[\frac{1}{2e_1}\big(\dot y_A\dot y^A-\frac{M}{R^2}
e_1^2\big)+\frac{1}{2e_2}\bigg(\Big(\frac{\dot b_A\dot b^A}{(\dot y_Ab^A)^2}
+\frac{1}{R^2}\Big)e_1^2-\delta\frac{e_2^2}{R^2}\bigg)\bigg]
\eqno{(23.a)}$$
where
$$
e_1\equiv\nu_5R^2, \qquad e_2\equiv -\nu_6R^2
\eqno{(23.b)}$$
and holonomic extra-constraints $T_{1,2,3}$ are imposed on $y^A$,
$b^A$.

This action is manifestly AdS invariant. What is more, it possesses
three local symmetries corresponding to three constraints depending on
momenta: $T_4$, $T_5$, $T_6$. They are
$$
\delta_5y^A=\bigg(\dot y^A-\frac{e_1^3}{e_2(\dot y_Ab^A)}\dot b^B\big(
\eta_{BC}+\frac{y_By_C}{R^2}\big)\dot b^Cb^A\bigg)m_5,
\eqno{(24.a)}$$
$$
\delta_5b^A=\frac{\dot y_Bb^B}{R^2}y^Am_5, \quad \delta_5e_1=\frac{\rm
d}{{\rm d}\tau}(e_1m_5), \quad \delta_5e_2=0;
\eqno{(24.b)}$$
$$
\delta_6y^A=\frac{e_1^3}{e_2(\dot y_Ab^A)}\dot b^B\big(\eta_{BC}+
\frac{y_By_C}{R^2}\big)\dot b^Cb^Am_6,
\eqno{(25.a)}$$
$$
\delta_6b^A=\big(\dot b^A-\frac{\dot y_Bb^B}{R^2}y^A\big)m_6, \quad
\delta_6e_2=\frac{\rm d}{{\rm d}\tau}(e_2m_6), \quad \delta_6e_1=0;
\eqno{(25.b)}$$
$$
\delta_4b^A=b^Am_4, \qquad \delta_4(\mbox{the rest variables})=0.
\eqno{(26)}$$
For instance, one can easily extract reparametrizations by taking
$m_5=m_6$ $=\mu$ and $m_4=0$. The third symmetry (26) reduces the theory
configuration manifold to $\cM$ while transformations (24), (25) are
the model's characteristic features those provide the Casimir operators
to conserve identically (see Eq. (19)).

Now let us briefly discuss the question about physical observables of the
theory. It is easily comprehended that all nontrivial physical
observables are functions of the Hamiltonian generators of the AdS
group modulo constraints. Indeed, all AdS group generators, obviously,
commute with the first-class constraints, on the other hand these
constraints reduce the original 12-dimensional phase space of the model
(if auxiliary constraints are taken into account) to the 8-dimensional
physical one. Consequently, physical subspace can be covariantly parametrized
with 10 generators of the AdS group subject to the two constraints.

\section{Reformulation of the model in terms of inner $\cM$ geometry}

In this Section, we give the another form for the action (23.a) which
could be treated as ``minimal covariant extension'' of the massive
spinning particle action in Minkowski space proposed earlier. Let us
consider some facts concerning $\cM$ geometry in order to expose this
formulation.

Let $x^m$ ($m=0,1,2,3$) be the local coordinates on the
surface (1). The induced metric is
$$
g_{mn}(x)=\eta_{AB}\frac{\pa y^A}{\pa x^m}~\frac{\pa y^B}{\pa x^n},
\eqno{(27)}$$
$$
\eta_{AB}{\rm d}y^A{\rm d}y^B = g_{mn}{\rm d}x^m{\rm d}x^n.
$$
The following 1-form of vierbein is associated with the metric (27):
$$
e_{ma}(x)=\frac{\pa y^A}{\pa x^m}\cF_{Aa}(y(x))=-y^A\frac{\pa\cF_{Aa}}
{\pa x^m},
\eqno{(28.a)}$$
where
$$
{\cF^A}_B(y)\equiv {(\cG^{-1})^A}_B(y)=\eta^{AC}\eta_{BD}{\cG^D}_C
\equiv{\cG_B}^A(y).
\eqno{(28.b)}$$
It is worth noting that
$$
{\cF^A}_5(y)\equiv{\cG_5}^A(y)=\frac{1}{R}y^A
\eqno{(29)}$$
as it follows from the very definition (6). Using the last formula it
is easy to check that $e_{ma}$ is really a vierbein, i.e.
$$
g_{mn}=e_{ma}e_{nb}\eta^{ab}.
\eqno{(30)}$$
The Lorentz connection associated with the vierbein (28.a) is
$$
{\omega_m}^{ab}(x)=\cF^{Aa}\pa_m{\cF_A}^b.
\eqno{(31)}$$
To verify this assertion it is enough to examine that the torsion
constructed on the base of Eqs. (28.a) and (31) vanishes:
$$
T^a_{nm}=\pa_n{e_m}^a-\pa_m{e_n}^a-{e_n}^b{\omega_{mb}}^a+{e_m}^b
\omega_{nba}=0.
\eqno{(32)}$$
Indeed,
$$
\pa_{[n}{e_{m]}}^a=\pa_{[n}y^A\pa_{m]}{\cF_A}^a
\eqno{(33.a)}$$
and
$$
{e_{[n}}^b{\omega_{m]b}}^a=\pa_{[n}y^A{\cF_A}^b{\cF^B}_b\pa_{m]}{\cF_B}^a=\qquad
$$
$$
=\pa_ny^A\pa_m{\cF_B}^A({\delta_A}^B-{\cF_A}^5{\cF^B}_5)=
\pa_{[n}y^A\pa_{m]}{\cF_A}^a,
\eqno{(33.b)}$$
since
$$
(\pa_ny^A){\cF_A}^5=\frac{1}{R}(\pa_ny^A)y_A=0.
\eqno{(33.c)}$$

Now let us consider the spinning part of $\cM$ -- two sphere
$S^2$.\footnote{\normalsize{Our two-component spinor notations mainly
coincide with those adopted in Ref. [8], except we number spinor indices
by values 0, 1 and define spinning matrices $\sigma_{ab}$ and
$\tilde\sigma_{ab}$ with additional minus sign in comparison with Ref. [8].}}
It is covered by the two charts, $z$ and $w$ are the complex coordinates
in these charts, and
$$
z=-\frac{1}{w}
\eqno{(34)}$$
in the overlap of charts.

The Lorentz group $SO^{\uparrow}(3,1)=SL(2,\cC)/\pm 1$ acts on $S^2$ by
means of fractional-linear transformations:
$$
z' = \frac{az-b}{-cz+d}, \qquad \|N\|={N_\alpha}^\beta=\left(
\begin{array}{cc} a & b\\ c & d\end{array}\right)\in SL(2,\cC).
\eqno{(35)}$$
It means that the two-component object
$$
z^\alpha\equiv (z)^\alpha =(1,z), \qquad \alpha =0,1
\eqno{(36)}$$
is transformed simultaneously as left Weyl spinor and spinor field on
$S^2$ under the Lorentz group (36):
$$
{z'}^\alpha =\left(\frac{\pa z'}{\pa z}\right)^{1/2} z^\beta
(N^{-1})_\beta{}^\alpha.
\eqno{(37)}$$
Let $p^a$ be a time-like 4-vector,
$$
p^2=p^ap_a<0.
\eqno{(38)}$$
One can associate with $p^a$ a smooth positive definite metric on $S^2$
of the form
$$
{\rm d}s^2 = \frac{4{\rm d}z{\rm d}{\bar z}}{(p^a\xi_a)^2}.
\eqno{(39.a)}$$
where
$$
\xi_a\equiv(\sigma_a)_{\alpha\dot\alpha}z^\alpha{\bar z}^{\dot\alpha}
\Rightarrow p^a\xi_a=p_{\alpha\dot\alpha}z^\alpha{\bar z}^{\dot\alpha},
\eqno{(39.b)}$$
$$
\xi_a\xi^a=0.
\eqno{(39.c)}$$
Metric (39.a) is Lorentz invariant in the following sense:
$$
\frac{{\rm d}z'{\rm d}{\bar z'}}{(p'_{\alpha\dot\alpha}{z'}^\alpha{\bar
z'}{}^{\dot\alpha})^2}=\frac{{\rm d}z{\rm d}{\bar z}}
{(p_{\alpha\dot\alpha}z^\alpha{\bar z}^{\dot\alpha})^2}
\eqno{(40.a)}$$
where
$$
p'_{\alpha\dot\alpha}={N_\alpha}^\beta{\bar N_\alpha}^{\dot\beta}
p_{\beta\dot\beta}.
\eqno{(40.b)}$$

In the case of massive spinning particle on the flat space, there
exists the only natural candidate to the role of $p^a$: it is tangent
vector to a particle's world line $\dx^a$. That is why one can
construct the following world line Lorentz-invariant
$$
\frac{4\dz\dbz}{(\dx^a\xi_a)^2}
\eqno{(41)}$$
which together with $\dx_a\dx^a$ constitute the set of building
blocks for the Lagrangian of massive spinning particle on the Minkowski
space [1]:
$$
S'=\int{\rm
d}\tau\left\{\frac{1}{2e_1}(\dx_a\dx^a-(me_1)^2)+\frac{1}{2e_2}\Big(
\frac{4\dz\dbz}{(\dx^a\xi_a)^2}+(\Delta e_2)^2\Big)\right\}.
\eqno{(42)}$$
This Lagrangian is invariant under global Poincar\'e transformations
when Poin\-car\'e-translations act trivially on $S^2$, and Lorentz group
is identified with diagonal of $SO(3,1)\big|{}_{R^{3,1}}\times
SO(3,1)\big|{}_{S^2}$, in accordance with Eq. (40).

We now show that the action of spinning particle on anti-de Sitter
space (23.a) could be derived by the minimal covariantization of
(23.a), i.e. by generalizing (23.a) to the form consistent with
general coordinate and local Lorentz covariance. If Lorentz
transformations are local, i.e. depend on $x^m$, Eq. (39.a) will not be
invariant because d$z$ is not local Lorentz-covariant differential.

However, the object
$$
Dz={\rm d}z +\frac{1}{2}{\rm d}x^m\omega_{mab}(x)\Sigma^{ab}
\eqno{(43.a)}$$
where
$$
\Sigma^{ab}\equiv(\sigma^{ab})_{\alpha\beta}z^\alpha z^\beta
\eqno{(43.b)}$$
is local Lorentz covariant:
$$
(Dz)'={\rm d}z'+\frac{1}{2}{\rm d}x^m\omega'_{mab}{\Sigma'}^{ab}=
\left(\frac{\pa z'}{\pa z}\right)Dz,
\eqno{(44.a)}$$
where
$$
\omega'_{mab}={\La_a}^c{\La_b}^d\omega_{mcd}+{\La_a}^c\pa_m\La_{bc},
\eqno{(44.b)}$$
(${\La_a}^c(x)$ are the local Lorentz transformations parameters) is
the transformation law for Lorentz connection easy derivable from Eqs.
(9.a) and (31). Taking the proper account of the relation (43) we find
that local Lorentz and general coordinate covariant generalization of
the ``flat'' action (42) is
$$
S_2=\int{\rm d}\tau\bigg\{\frac{1}{2e_1}(g_{mn}\dx^m\dx^n-(me_1)^2)+
\frac{1}{2e_2}\bigg(\frac{4|Dz/{\rm d}\tau|^2}{(\dx{}^me_{ma}\xi^a)^2}+
(\Delta e_2)^2\bigg)\bigg\}.
\eqno{(45)}$$
It turns out that the following equality takes place:
$$
4\frac{|Dz/{\rm d}\tau|^2}{(\dx{}^me_{ma}\xi^a)^2}=\frac{\dot b_A\dot b{}^A}
{(\dot y_Ab^A)^2}+\frac{1}{R^2}
\eqno{(46)}$$
where the following parametrization of the light-cone (7) is used
$$
u^a=\xi^aE(u)
\eqno{(47)}$$
and $E(u)$ is some function on the light-cone.

To prove the equality (46) one need to employ the properties of
${\cF^A}_a$ (5), (28.a), (29) and definitions of ${\omega_m}^{ab}$ and
$e_{ma}$ through ${\cF^A}_a$ (28.a), (31). Then one can make sure that
$$
\dot y_Ab^A=\dx^me_{ma}\xi^aE,
\eqno{(48.a)}$$
$$
{\cF^A}_a\dF_{Ab}=\dx^m\omega_{mab}.
\eqno{(48.b)}$$
The useful identity
$$
\varepsilon^{\alpha\beta}=z^\alpha\pa_zz^\beta -z^\beta\pa_zz^\alpha
\eqno{(49)}$$
allows one to prove that
$$
\dxi_a\dxi^a=4\dz\dbz,
\eqno{(50.a)}$$
$$
\dxi^a\xi^b-\xi^a\dxi^b=2((\sigma^{ab})_{\alpha\beta}z^\alpha
z^\beta\dbz -(\tilde\sigma^{ab})_{\dot\alpha\dot\beta}{\bar
z}^{\dot\alpha}{\bar z}^{\dot\beta}\dz ).
\eqno{(50.b)}$$
Using (48)--(50) one obtains
$$
\dxi^2+{\cF^A}_a\dF_{Ab}(\dxi^a\xi^b-\xi^a\dxi^b)=\qquad\qquad
$$
$$
=4(\dz\dbz +\frac{1}{2}\dx^m\omega_{mab}((\sigma^{ab})_{\alpha\beta}
z^\alpha z^\beta\dbz -(\tilde\sigma^{ab})_{\dot\alpha\dot\beta}{\bar
z}^{\dot\alpha}{\bar z}^{\dot\beta}\dz )).
\eqno{(51.a)}$$
and
$$
({\cF^A}_a\dF_{Ab}\xi^a\xi^b)R^2+(\dx^me_{ma}\xi^a)^2=|\dx^m
\omega_{mab}(\sigma^{ab})_{\alpha\beta} z^\alpha z^\beta|^2R^2.
\eqno{(51.b)}$$
Two last equalities are directly equivalent to Eq. (46). Thus,
$$
S_1=S_2
\eqno{(52.a)}$$
under the identification (46) and
$$
m^2=M/R^2, \qquad \Delta^2=-\delta/R^2.
\eqno{(52.b)}$$

So, we have two formulations for a given spinning particle on AdS
background, $S_1$ and $S_2$. The first formulation exhibits AdS
invariance of the model in the straightforward way, while the second
one describes theory in terms of inner $\cM$ geometry without
introduction of auxiliary degrees of freedom. It might be well to
mention that the derivation procedure used in Sec. 3 can be
successfully performed in the inner $\cM$ terms. Namely let us consider
the space which is the cotangent bundle to ${\cal M}^6$
$$
\{x^m,p_n\}=\delta^m_n, \qquad \{z,p_z\}=1, \qquad \{\bar z,p_{\bar
z}\}=1.
\eqno{(53)}$$
The classical counterparts of AdS -- Casimir operators are
$$
C_1=R^2g^{mn}\nabla_m\nabla_n,
\eqno{(54.a)}$$
$$
C_2=\frac{1}{2}(C_1)^2+R^2(\nabla^me_{ma}\cF^a)^2,
\eqno{(54.b)}$$
where
$$
\nabla_m\equiv p_m-\frac{1}{2}{\omega_m}^{ab}(x)S_{ab},
\eqno{(55.a)}$$
$$
S_{ab}=-(\sigma_{ab})_{\alpha\beta}z^\alpha z^\beta p_z+
(\tilde\sigma_{ab})_{\dot\alpha\dot\beta}{\bar z}^{\dot\alpha}{\bar
z}^{\dot\beta}p_{bar z},
\eqno{(55.b)}$$
$$
\cF^a=2(\sigma^a)_{\alpha\dot\beta}z^\alpha{\bar z}^{\dot\beta}|p_z|.
\eqno{(55.c)}$$
Due to the general coordinate and local Lorentz covariance, relations
(54.a,b) can be verified in any useful coordinate system, for example,
in those for which AdS generators look like
$$
L_{ab}=x_ap_b-x_bp_a+S_{ab},
\eqno{(56.a)}$$
$$
L_{a5}=Rp_a+\frac{1}{4R}(2x_ax^bp_b-x^2p_a)+\frac{1}{2R}x^bS_{ab},
\eqno{(56.b)}$$
while the tetrade reads as follows
$$
e_{ma}=\eta_{ma}\left(1+\frac{x^2}{4R^2}\right)^{-1},
\eqno{(57.a)}$$
$$
{\omega_m}^{ab}=-\frac{1}{2R^2}(x^a{\delta_m}^b-x^b{\delta_m}^a)
\left(1+\frac{x^2}{4R^2}\right)^{-1}.
\eqno{(57.b)}$$
Now, the following action
$$
S_2^{\rm H}=\int{\rm d}\tau\bigg\{p_m\dx^m+p_z\dz +p_{\bar z}\dbz-
\frac{e_1}{2}\bigg(\frac{C_1}{R^2}+m^2\bigg)-\qquad
$$
$$
\qquad-\frac{e_2}{2}\bigg(\frac{1}{R^2}\bigg(C_2-\frac{1}{2}C_1^2
\bigg)-\Delta^2\bigg)\bigg\}
\eqno{(58)}$$
is nothing but Hamiltonian formulation of $S_2$, i.e. if one exclude
momenta $p_m$, $p_z$, $p_{\bar z}$ by making use of their equations of
motion, he arrive exactly to $S_2$.

The following remark is very much to the point: the second formulation
(45), (54), (58) seems to be well defined on a general curved space,
i.e. when $g_{mn}(x)$ is arbitrary.

It turns out, however, that the classical dynamics of the model is
non-contradictory on maximal symmetric spaces (i.e., de
Sitter--Minkowski--Anti de Sitter) only. It can be seen as follows. The
action functional is obviously reparametrization invariant on general
background (see also Eq. (61)), that's why a first class constraint has
to exist in Hamiltonian formulation. At the same time, Hamiltonian
formulation (54), (55) and (58) is determined by two constraints (54)
subject to the following Poisson bracket relation:
$$
\{C_1,C_2\}=R^4{R_{nm}}^{cd}e^{am}\cF_aS_{cd}\nabla^n
\eqno{(59)}$$
where ${R_{nm}}^{cd}$ is Riemann tensor.

The identity
$$
S^{cd}\cF_d\equiv 0
\eqno{(60)}$$
makes the commutator (59) to be zero if and only if ($acd$)-traceless
projection of ${R_{nm}}^{cd}e^{em}$ vanishes. Making use of the Bianchi
identities one can see that the space-time has the constant curvature.
Otherwise the constraints $C_1$, $C_2$ turn out to be of the second
class that contradicts to the reparametrization invariance of the
action.

Unfortunately, it still remains unclear whether it is possible to find
appropriate curvature depending contributions to constraints $C_1$,
$C_2$ to make them involutive.

To conclude this section let us note that Lagrange multipliers $e_1$
and $e_2$ can also be eliminated with the aid of their equations of
motion. The result is ``Nambu--Goto form'' of the action
$$
S=\int{\rm d}\tau\sqrt{g_{mn}\dx^m\dx^n\left(m^2-4\Delta\frac
{|Dz/{\rm d}\tau|}{|\dx{}^me_{ma}\xi^a|}\right)}=
$$
$$
=\int{\rm d}\tau\sqrt{\dot y_A\dot y^A\bigg(m^2-2\Delta\bigg(\frac
{\dot b_A\dot b^A}{(\dot y_Ab^A)^2}+\frac{1}{R^2}\bigg)^{1/2}\bigg)}.
\eqno{(61)}$$
For $\Delta=0$ this expression apparently reduces to the action of
spinless particle on AdS background, on the other hand the limit
$R\Rightarrow\infty $ results in the $(m,s)$-particle action [1] in Minkowski
space.

\section{Conclusion}

Let us give a brief overview of the results and some comments. We have
suggested the model of a spinning particle which propagates in $d=4$
Anti-de Sitter space. The configuration space of the theory is
six-dimensional manifold $\cM$ being the product of $d=4$ AdS space and
two-dimensional sphere. The values of the phase-space counterparts of
the AdS group Casimir operators are fixed by two abelian first-class
constraints to be arbitrary real numbers. The model can conceptually be
treated as an universal and minimal AdS spinning particle theory in the
sense that configuration manifold is spin-independent and has the
minimal possible dimension which still provides dynamical activity both
for particle position and spinning degree of freedom. As a consequence
of model's minimality property all the observables turn out to be
functions of the AdS group generators only. So, the model quantization
problem reduces to the construction of the irreducible unitary
representation of the AdS group with given quantum numbers. This
problem has been solved in Ref.[9] where irreducible representations
have been found in the case of bounded energy that provides a proper
particle interpretation.

It is pertinent to note that it is not evident how to construct an
explicit Hilbert space realization (e.g. coordinate one) for this
(``constrained'') quantum mechanics, i.e. the question is how to frame
the AdS (spin)tensor field on $\cM$ with an appropriate Hilbert space
structure. This problem has been exhaustively studied for the flat
space case in the paper [1] where all such realizations were
classified. We intend to give the similar investigation for the AdS
case in the forthcoming publication.

Let us mentioned that this model could be generalized for
$d$-dimensional ($d>4$) AdS space in the straightforward way (one need
to assume $y^A$, $b^A$ to be $d$-dimensional). However, such a simplest
extension would not able to describe the most general case of the
higher-dimensional AdS spinning particle. The reason is that the
higher-dimensional AdS group has some extra Casimir operators which
turn out to vanish identically in the cotangent bundle of ${\cal
M}^d_\rho\times S^{d-2}$. Thus the straightforward higher-dimensional
extension of the proposed model could describe only the particles
associated with the irreducible AdS group representations characterized
by zero eigenvalues of these extra Casimir operators.

\section*{Acknowledgments}

The researches of S.L.L., A.Yu.S., and A.A. Sh. are supported in part
by European Community Commission contract INTAS-93-2058 and by the
International Science Foundation grant No M2I000. Researches of S.M.K.
are supported by Alexander von Gumboldt Foundation fellowship.

\section*{References}

\noindent
[1] S.M. Kuzenko, S.L. Lyakhovich, and A.Yu. Segal, Preprint
TSU-TP-94-8, hep-th/9403196 ``A Geometric model of arbitrary spin
massive particle'', to appear in Int. J. Mod. Phys.{\bf A},(1994)\newline
[2] M.A. Vasiliev, Phys. Lett., {\bf B 243}, 378 (1990)\newline
[3] F.A. Berezin and M.S. Marinov, JETP Lett. {\bf 21}, 678 (1975);
Ann. Phys. {\bf 104}, 336 (1977).\newline
[4] A. Barducci, R. Casalbuoni, and L. Lusanna, Nuovo Cim. {\bf A35},
377 (1976).\newline
[5] L. Brink, S. Deser, B. Zumino, P. Di Vecchia, and P.S. Howe, Phys.
Lett. {\bf B 64}, 435 (1976).\newline
[6] P.S. Howe, S. Penati, M. Pernici, and P. Townsend, Phys. Lett. {\bf
B 215}, 255 (1988).\newline
[7] R. Marnelius and U. M\.{a}rtensson, Nucl. Phys. {\bf B335}, 395
(1991); Int. J. Mod. Phys. {\bf A6}, 807 (1991). U. M\.{a}rtensson,
Int. J. Mod. Phys. {\bf A8}, 5305 (1994).\newline
[8] J. Wess and J. Bagger, {\it Supersymmetry and Supergravity}
(Princeton Univ. Press., Princeton, 1983).\newline
[9] N.T. Evans, J. Math. Phys., {\bf 8}, 2 (1967).
\end{document}